\documentclass[12pt]{iopart}
\usepackage{iopams}
\usepackage{graphicx}
\usepackage{graphics}

\begin{document}

\newcommand{\vc}[1]{\mathbf{#1}}

\title{Dynamic charge inhomogenity in cuprate superconductors}
\author{Thomas Bauer, Claus Falter}
\ead{falter@uni-muenster.de}
\address{Institut f\"ur Festk\"orpertheorie, Westf\"alische Wilhelms-Universit\"at,\\
Wilhelm-Klemm-Str.~10, 48149 M\"unster, Germany}
\date{\today}

\begin{abstract}
The inelastic x-ray scattering spectrum for phonons of
$\Delta_{1}$-symmetry including the CuO bond-stretching phonon
dispersion is analyzed by a Lorentz fit in HgBa$_{2}$CuO$_{4}$ and
Bi$_{2}$Sr$_{2}$CuO$_{6}$, respectively, using recently calculated
phonon frequencies as input parameters. The resulting mode frequencies
of the fit are almost all in good agreement with the calculated data.
An exception is the second highest $\Delta_{1}$-branch compromising the
bond-stretching modes which disagrees in both compounds with the
calculations. This branch unlike the calculations shows an anomalous
softening with a minimum around the wavevector
$\vc{q}=\frac{2\pi}{a}(0.25,\,0,\,0)$. Such a disparity with the
calculated results, that are based on the assumption of an undisturbed
translation- and point group invariant electronic structure of the CuO
plane, indicates some {\it static} charge inhomogenities in the
measured probes. Most likely these will be charge stripes along the CuO
bonds which have the strongest coupling to certain longitudinal
bond-stretching modes that in turn selfconsistently induce
corresponding {\it dynamic} charge inhomogenities. The symmetry
breaking by the mix of dynamic and static charge inhomogenities can
lead to a reconstruction of the Fermi surface into small pockets.
\end{abstract}


\maketitle

In recent work \cite{Uchi04,Graf08} the IXS phonon spectrum of
optimally doped HgBa$_{2}$CuO$_{4}$ and Bi$_{2}$Sr$_{2}$CuO$_{6}$,
respectively, has been measured for the $\Delta_{1}$-symmetry modes.
The spectra have been interpreted by a Voigt fit function. For the
interesting high frequency part of the spectra where the CuO
bond-stretching modes (BSM) are expected that are known to display an
anomalous softening upon doping in the cuprates
\cite{Pint05,Pint98,Pint99,Reich96,Fuku05,Pint06,McQ01,d'Astu02,d'Astu03,Braden05},
only one phonon peak has been fitted in case of HgBa$_{2}$CuO$_{4}$
\cite{Uchi04} and two in case of Bi$_{2}$Sr$_{2}$CuO$_{6}$
\cite{Graf08}.

\begin{figure}%
 \includegraphics[]{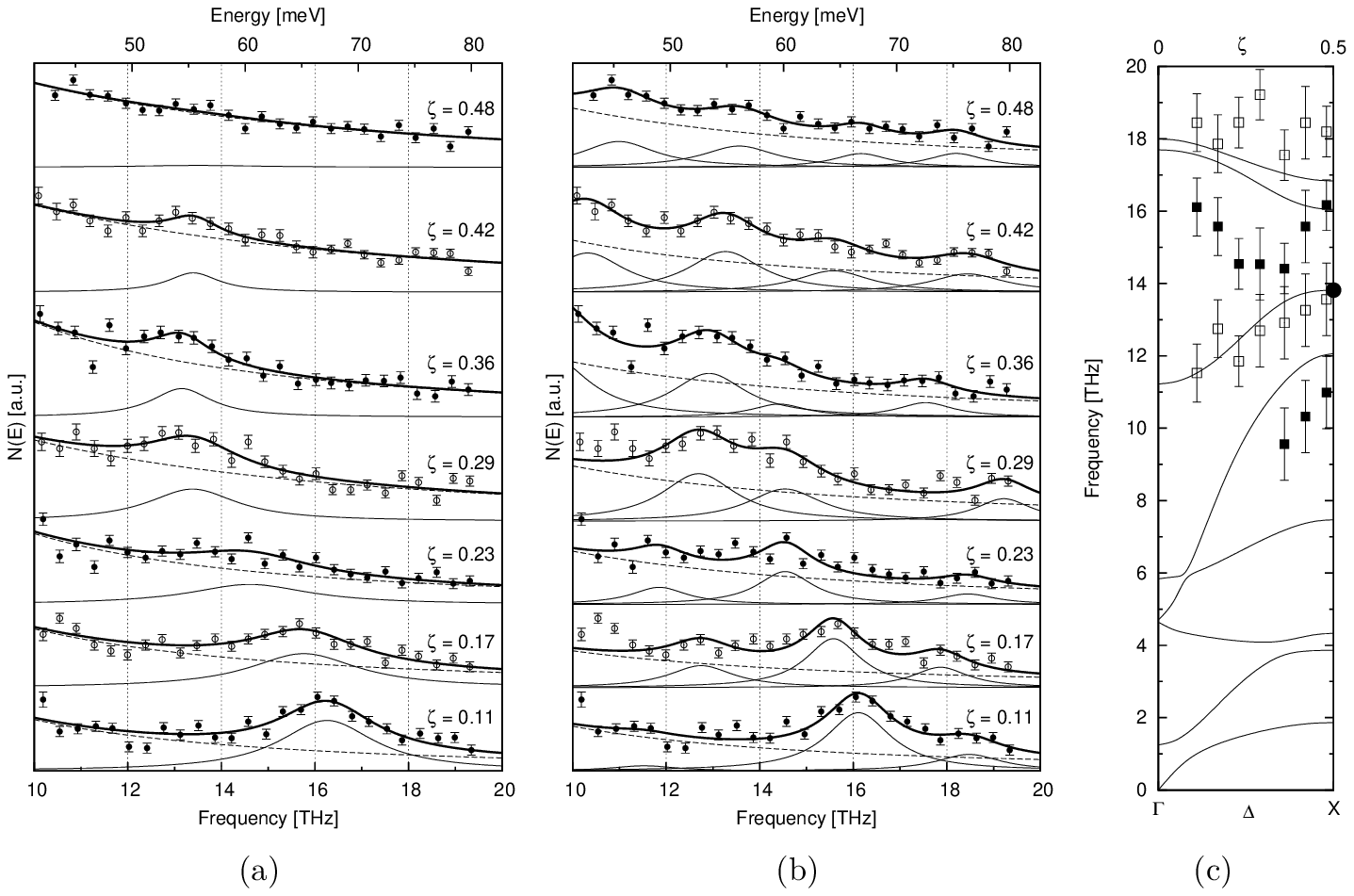}%
 \caption{(a) Reconstructed IXS phonon spectrum of HgBa$_{2}$CuO$_{4}$
 from \cite{Uchi04} for $\vc{q}=\frac{2\pi}{a}(\zeta,0,0)$ using
 a Lorentz fit with one mode only (thick line), the
 corresponding Lorentz peaks of the modes
 are also shown (thin line). The broken line represents the
 elastic background. The inelastic cross section $N(E)$ is given
 in arbitrary units. (b) Same as in (a) but using a Lorentz
 fit on the basis of the calculated
 phonon dispersion for modes of $\Delta_1$-symmetry in the energy
 range considered \cite{Bauer10}.
 (c) Calculated phonon dispersion of HgBa$_{2}$CuO$_{4}$ for modes
 of $\Delta_1$-symmetry according to \cite{Bauer10}.
 The symbols \fullsquare\ and \opensquare\ denote the phonon frequencies
 from the Lorentz fit and \fullcircle\ the half-breathing mode.
 The vertical bars indicate the FWHM of the peaks determined
 in the fitting.}\label{fig01}%
\end{figure}

On the other hand, from our microscopic calculation of the phonon
dynamics of HgBa$_{2}$CuO$_{4}$ and Bi$_{2}$Sr$_{2}$CuO$_{6}$
\cite{Bauer10} we find in the high energy region of the IXS spectra
along $\frac{2\pi}{a}(\zeta,0,0)$ up to four $\Delta_{1}$ modes as far
as HgBa$_{2}$CuO$_{4}$ is concerned and up to five $\Delta_{1}$ modes
in Bi$_{2}$Sr$_{2}$CuO$_{6}$. These modes interact strongly with each
other and constitute a highly nontrivial anticrossing scenario. As a
consequence a multiple mode fit is necessary to obtain more reliable
results. The strength of the measured peaks strongly decreases for
larger $\zeta$ values such that at $\zeta \approx 0.5$ the peaks are
very weak, see e.g. \fref{fig01}, and an unique interpretation taking
only one or two peaks into account is incomplete because of the many
modes of the same symmetry in this region of energy. Nevertheless, it
is very interesting that in the limited Voigt fit of the spectra in
\cite{Uchi04,Graf08} a hint to an anomalous softening of the
BSM-dispersion is extracted at $\zeta \approx 0.3$ in the Hg compound
and at $\zeta \approx 0.25$ in the Bi compound. Such an enhanced
softening at these $\zeta$ values has not been detected in our
calculations for the $\Delta_{1}$ modes. Only the presumably generic
softening in the cuprates of the half-breathing mode (HBM) in
HgBa$_{2}$CuO$_{4}$ and of two half-breathing modes in
Bi$_{2}$Sr$_{2}$CuO$_{6}$, generated by the strong nonlocal
electron-phonon interaction (EPI) mediated by the charge fluctuations
(CF) on the outer electron shells of the ions, has been found in our
calculations \cite{Bauer10}.

In the presence of these theoretical results and the complex
anticrossing of the $\Delta_{1}$ modes it is very desirable to perform
a more complete analysis of the IXS measurements on the basis of the
calculation of the dispersion of the $\Delta_{1}$ modes as given in
\cite{Bauer10}. In such a reconstruction of the IXS phonon spectra we
do not apply the more complicated Voigt fit but use instead a more
simple Lorentz fit function, which does the job very well. This can be
concluded from \fref{fig01} (a) where the IXS spectrum of
HgBa$_{2}$CuO$_{4}$ \cite{Uchi04} is reconstructed for
$\vc{q}=\frac{2\pi}{a}(\zeta,0,0)$ using as in \cite{Uchi04} only one
peak in the Lorentz fit. The phonon frequencies fitted at $\zeta =
0.11$, 0.17, 0.23, 0.29, 0.36, 0.42 and 0.48 are 16.28 (16.32), 15.76
(15.73), 14.61 (14.65), 13.38 (13.73), 13.15 (13.14), 13.40 (13.53) and
13.51 (13.63) THz. The data within the brackets are the frequencies as
obtained from the Voigt fit \cite{Uchi04}. We recognize a very good
agreement between both fitting procedures. As far as
Bi$_{2}$Sr$_{2}$CuO$_{6}$ is concerned we also get nearly
undistinguishable results for a two mode fit in the two fitting
schemes, so we take the more simple Lorentz fit in this work.

In \fref{fig01} (b) we present the results of the Lorentz fit for the
analysis the IXS spectrum of HgBa$_{2}$CuO$_{4}$ within the framework
of the calculated $\Delta_{1}$-modes in the relevant energy range
\cite{Bauer10}. As already mentioned up to four mode frequencies must
be considered. The calculated frequencies at the corresponding $\zeta$
values are taken as input parameters of the Lorentz fit. From the
results displayed in \fref{fig01} (b) we find that the IXS spectrum is
well represented by the multiple mode fit. The Lorentz peaks as
obtained by the fit are also shown in the figure for each $\zeta$
value.

The resulting phonon frequencies from the fit are shown in \fref{fig01}
(c) together with the calculated dispersion of the $\Delta_{1}$ modes
\cite{Bauer10}. With exception of the second highest $\Delta_{1}$
branch comprising the BSM that displays the anomalous softening
starting at $\zeta \approx \frac{1}{8}$ with a minimum around $\zeta =
\frac{1}{4}$, the resulting mode frequencies are in reasonable
agreement with the calculated branches.

\begin{figure}%
 \includegraphics[]{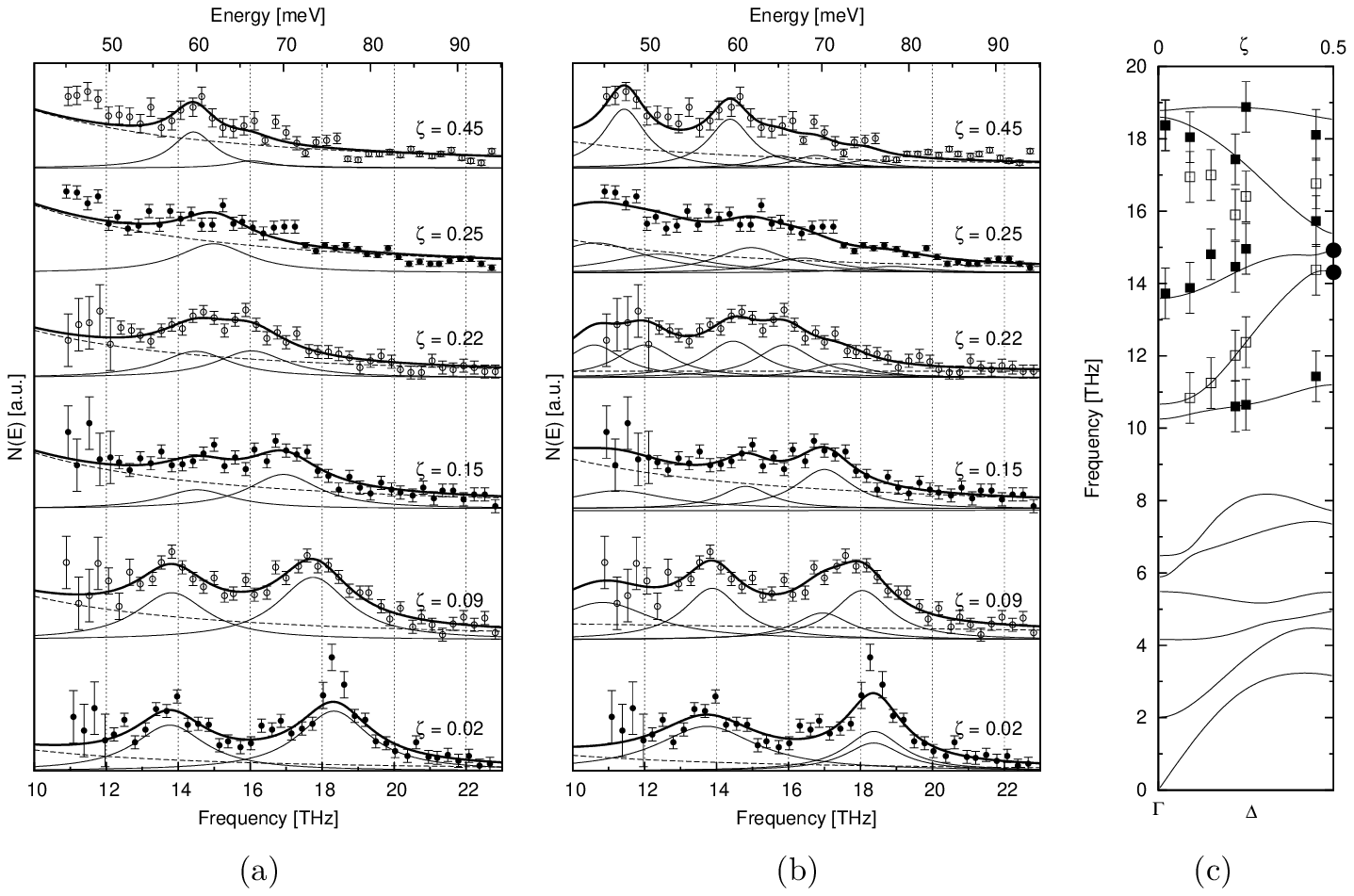}%
 \caption{(a) Reconstructed IXS phonon spectrum of Bi$_{2}$Sr$_{2}$CuO$_{6}$
 from \cite{Graf08} for $\vc{q}=\frac{2\pi}{a}(\zeta,0,0)$ using
 a Lorentz fit with two modes only (thick line), the
 corresponding Lorentz peaks of the modes
 are also shown (thin line). The broken line represents the
 elastic background. The inelastic cross section $N(E)$ is given
 in arbitrary units. (b) Same as in (a)
 but using a Lorentz fit on the basis of the calculated
 phonon dispersion for modes of $\Delta_1$-symmetry in the energy
 range considered \cite{Bauer10}.
 (c) Calculated phonon dispersion of Bi$_{2}$Sr$_{2}$Cu$_{6}$ for modes
 of $\Delta_1$-symmetry according to \cite{Bauer10}.
 The symbols \fullsquare\ and \opensquare\ denote the phonon frequencies
 from the Lorentz fit and \fullcircle\ the half-breathing modes.
 The vertical bars indicate the FWHM of the peaks determined
 in the fitting.}\label{fig02}%
\end{figure}%

In case of Bi$_{2}$Sr$_{2}$CuO$_{6}$ the Lorentz fit of the IXS
spectrum on the basis of the calculated $\Delta_{1}$-modes
\cite{Bauer10} is shown in \fref{fig02} (b) and likewise as in
HgBa$_{2}$CuO$_{4}$ a good resolution of the spectrum is achieved. The
extracted phonon frequencies from the fit are shown in \fref{fig02}
(c). For a comparison the calculated dispersion of the
$\Delta_{1}$-modes \cite{Bauer10} are displayed in the figure. Leaving
aside the second highest branch we find a good agreement of the
calculated branches with the modes from the Lorentz fit. As for
HgBa$_{2}$CuO$_{4}$ the second highest $\Delta_{1}$ branch displays an
anomalous softening starting at $\zeta \approx \frac{1}{8}$ with a
minimum around $\zeta = \frac{1}{4}$.

\begin{figure}%
 \includegraphics[]{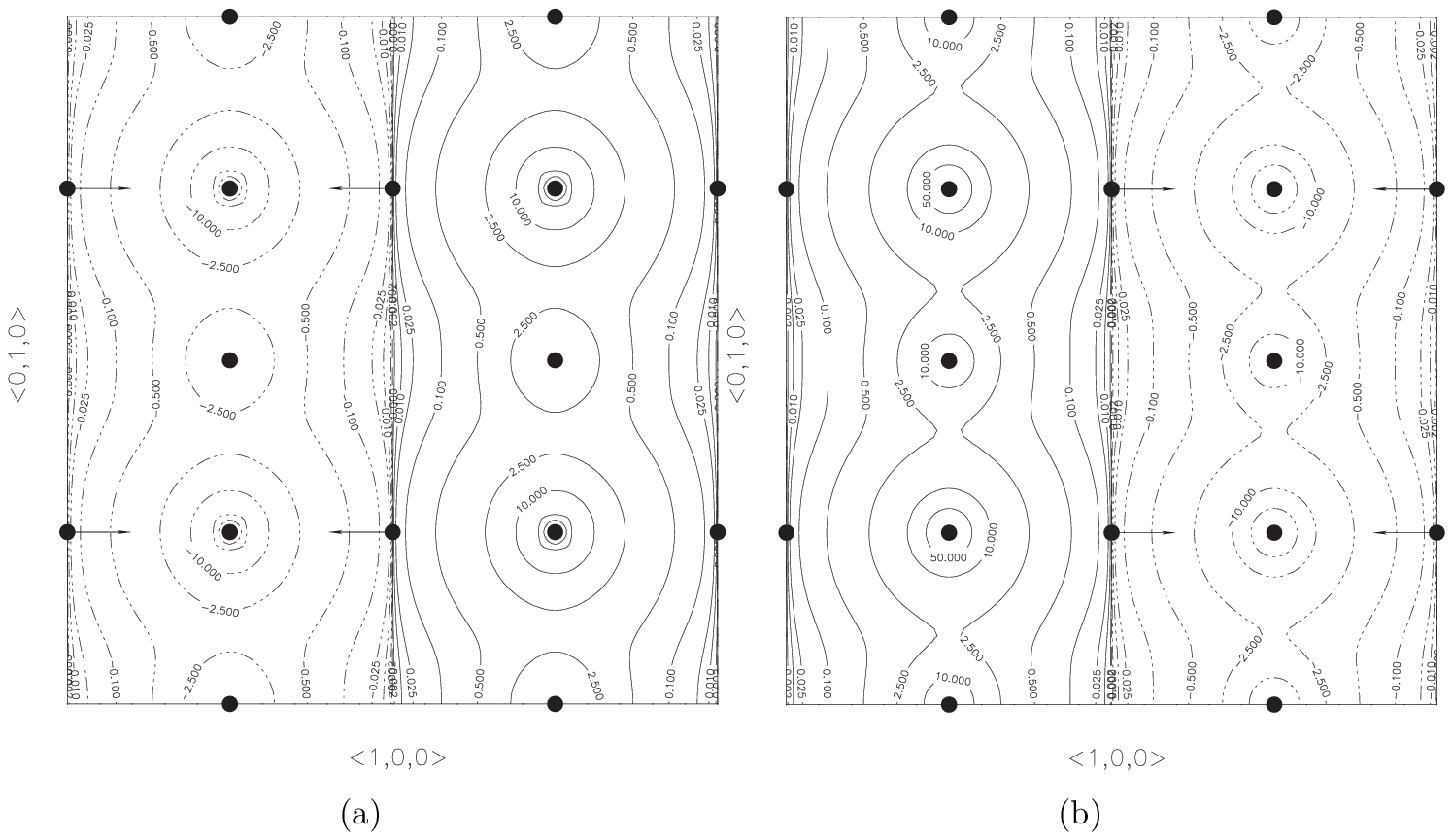}%
 \caption{Dynamical charge stripes $\delta\rho$ as induced by the half-breathing mode
 (with the lower frequency) in Bi$_{2}$Sr$_{2}$CuO$_{6}$ (a) and
 in HgBa$_{2}$CuO$_{4}$ (b). The arrows indicate the vibrating oxygen ions.
 Units are in $10^{-4}$ $e^2/a_{\rm B}$. Regions of space where electrons are
 accumulated are characterized by full lines and those where electrons are
 pushed away by broken-dotted lines.}\label{fig03}%
\end{figure}%

Strong coupling of the BSM with dynamic charge stripes generated
selfconsistently via strong nonlocal electron-phonon interaction (EPI)
by the BSM themselves leading to an anomalous softening has already
been found earlier in the cuprate based superconductors for the case of
an undisturbed translation- and point group invariant electronic
structure of the CuO plane without any charge inhomogenities
\cite{Falter05,Falter06,Falter97,Bauer08,Falter00,Bauer09a}. The same
holds true for HgBa$_{2}$CuO$_{4}$ and Bi$_{2}$Sr$_{2}$CuO$_{6}$ as can
be seen from the calculations in this work displayed in \fref{fig03},
where the charge redistribution $\delta\rho(\vc{r})$ for the
half-breathing mode (HBM) $(\zeta = \frac{1}{2}$) induced by nonlocal
EPI is shown. In the HBM propagating in $\Delta \sim (1,0,0)$ direction
the O$_{x}$ ions move coherently along the CuO bond towards or away
from the silent Cu ion. For the detailed displacement pattern, see
\cite{Bauer10}.

\Fref{fig03} illustrates that the HBM creates in both compounds dynamic
charge ordering $\delta\rho$ via charge fluctuations at the ions due to
nonlocal EPI in form of localized stripes of alternating sign in the
CuO plane. The period of the pattern is twice the lattice constant for
$\zeta = \frac{1}{2}$. In case of Bi$_{2}$Sr$_{2}$CuO$_{6}$ these
dynamic charge inhomogenities are only shown for one of the two HBM
modes found in the calculations namely the one with the lower
frequency, see \fref{fig02} (c), because the charge redistribution is
very similar for the other HBM. For HgBa$_{2}$CuO$_{4}$ we obtain only
one HBM, see \fref{fig01} (c), which induces the dynamic stripe pattern
in \fref{fig03} (b). In the HBM the dynamic charge stripes point along
the $x$- or $y$-axis, respectively, if the oxygen ions move along the
CuO bond in $y$- or $x$-direction, respectively.

It should be noted that for the oxygen breathing mode at the $X$ point,
see \cite{Bauer10}, the dynamic stripes excited point along the
diagonals of the CuO plane. Furthermore, it should be remarked that
dynamic charge inhomogenities like the dynamic stripe patterns would
not be present in systems with only {\it local} EPI at work. For
example in a high density homogeneous electron gas prevailing in simple
metals. This is because of perfect screening of the changes of the
Coulomb potential related to the displacement of the ions in a phonon
mode. On the other hand, our calculations exhibit a strong nonlocal EPI
in the cuprates essentially due to the poor screening in these
materials with a large component of ionic binding, see. e.g.
\cite{Bauer09a}. This fact demonstrates the special role of a strong
nonlocal EPI for the physics of these compounds in their normal as well
as in the superconducting state.

\begin{figure}%
 \includegraphics[width=0.8\linewidth]{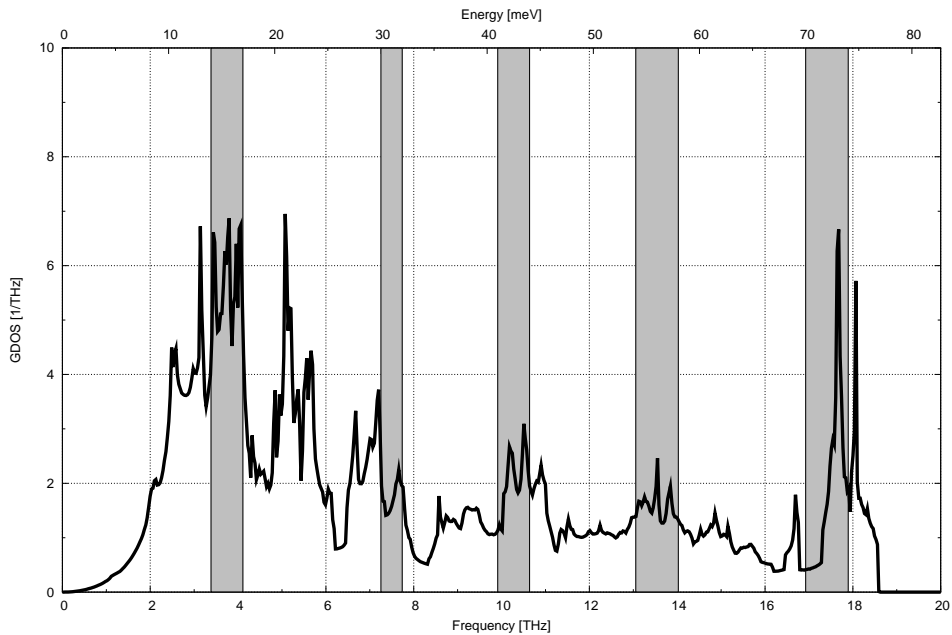}%
 \caption{Calculated phonon density of states
 for Bi$_{2}$Sr$_{2}$CuO$_{6}$. The gray domains mark the
 peaks of the Eliashberg spectral function \cite{Zhao10}.}\label{fig04}%
\end{figure}%

Such a strong coupling is also underlined by recent experimental
results for the quasi-particle dispersion in
(Bi,Pb)$_2$Sr$_2$CuO$_{6+\delta}$ \cite{Zhao10}. The super-high
resolution laser-based angle-resolved photoemission spectroscopy
measurements reported in \cite{Zhao10} show unambiguously that the $\sim 70$ meV
nodal kink in the dispersion is due to strong EPI with multiple phonon
modes as seen in the Eliashberg spectral function extracted from the
measured real part of the electron self-energy. Remarkably, the peak
structure of the Eliashberg function is in very good agreement with our
calculated phonon density of states for Bi$_{2}$Sr$_{2}$CuO$_{6}$
\cite{Bauer10}, see \fref{fig04}.

Quite generally, our calculations for the cuprates demonstrate the
enhanced susceptibility of the BSM for inducing dynamic charge
inhomogenities in form of stripes in the CuO plane. So, it can safely
argued that the latter also will occur in a modified way in systems
with static stripe charge order breaking translation- and point group
invariance. The anomalous softening around $\zeta = \frac{1}{4}$ found
by the analysis of the IXS phonon spectra then reflects the strong
nonlocal coupling of the BSM with their selfconsistently induced
dynamic charge inhomogenities (dynamic stripes) but now in probes with
a static charge inhomogenity (static stripes) presumably present. Excess
holes in a stripe will even enhance the strong nonlocal coupling to the
BSM and consequently also the magnitudes of the CF's and the related
mode softening is increased. An interrelation between the number of
holes and softening is supported by the experiment and our calculations
for the Cu-O-bond stretching modes \cite{Falter05,Falter06}.

The symmetry breaking of the electronic structure by charge
inhomogenity in form of static stripe patterns accompanied by
phonon-induced dynamic fluctuating patterns shown by our calculations
to occur very likely may lead to a corresponding reconstruction of the
Fermi surface into small pockets. This reconstruction is in turn the
origin for the quantum oscillations measured in some cuprates, see e.g.
\cite{Doiron07}. In this context it is interesting, that in
\cite{Taillefen09} the quantum oscillations are attributed to symmetry
breaking by a certain type of stripe order in form of a combined
charge/ spin modulation. Other authors attribute the small Fermi
surface pockets to magnetic fluctuations \cite{Harrison09}, to
$d$-density wave \cite{Chakravarty08} or to an antiferromagnetic
$(\pi,\pi)$ spin density wave \cite{Lin05}.

Finally, it should be added that in the literature dynamic charge
inhomogenities related to electron-phonon coupling are discussed from
the experimental side for La$_{2-x}$Ba$_{x}$CuO$_{4}$,
YBa$_{2}$Cu$_{3}$O$_{7}$ or Ba$_{0.6}$K$_{0.4}$BiO$_{3}$ in
\cite{Reznik06,d'Astuto08,Pint04,McQ99,Pint03,Chung03,Braden01,Braden02}
and from the theoretical side e.g. in \cite{Kaneshita02}.

In summary, our analysis of the IXS phonon spectra for
HgBa$_{2}$CuO$_{4}$ and Bi$_{2}$Sr$_{2}$CuO$_{6}$ within the framework
of calculated phonon dispersion curves points to the existence of
dynamic charge inhomogenities generated by the BSM via strong nonlocal
EPI of the CF-type. These modes excite dynamic charge order and
consequently can also serve as indirect probes for corresponding static
charge order in the cuprates. Thus, we have a mix of static and
fluctuating charge order in the materials.

\end{document}